\def\boxit#1#2{\hbox{\vrule  
                        \vbox{\hrule \vskip #1
                             \hbox{\hskip #1
                                 \vbox{\hsize=#1}#2%
                              \hskip #1}%
                         \vskip #1 \hrule}%
                      \vrule}}

\def\half{\frac{1}{2}}
\def\Ai{{\rm Ai}}
\documentclass{elsarta}
\usepackage{graphicx}
\begin{document}
\hspace*{2.5 in}CUQM-119, HEPHY-PUB 822/06\hfil\break
\hspace*{2.5 in}math-ph/0702047
\vspace*{0.4 in}

\begin{frontmatter}
\title{Schr\"odinger secant lower bounds to semirelativistic eigenvalues}\author{Richard L. Hall$^{1}$ and
Wolfgang Lucha$^{2}$}
\address{$^{1}$Department of Mathematics and
Statistics, Concordia University,\\1455 de Maisonneuve Boulevard
West, Montr\'eal, Qu\'ebec, Canada H3G 1M8\\E-mail:
rhall@mathstat.concordia.ca}
\address{$^{2}$Institute for High Energy Physics, Austrian Academy of
Sciences,\\Nikolsdorfergasse 18, A-1050 Vienna, Austria\\E-mail:
wolfgang.lucha@oeaw.ac.at}
\maketitle\markboth{Richard L.~Hall and
Wolfgang Lucha}{Schr\"odinger secant lower bounds $\dots$}
\begin{abstract}It is shown that the ground-state eigenvalue of a
semirelativistic Hamiltonian of the form $H = \sqrt{m^2+p^2} + V$
is bounded below by the Schr\"odinger operator $m + \beta p^2 +
V,$ for suitable $\beta >0.$ An example is
discussed.\end{abstract}
\end{frontmatter}

\section{Introduction}We study semirelativistic Hamiltonians $H$
composed of the relativistically correct expression $K(p^2) =
\sqrt{m^2 + p^2},$ $p \equiv |{\bf p}|,$ for the energy of a free
particle of mass $m$ and momentum ${\bf p},$ and of a
coordinate-dependent static interaction potential $V(r),$ $r
\equiv |{\bf r}|,$ which may be chosen arbitrarily, apart from the
constraint imposed on $H$ that it be bounded from below: $$H =
\sqrt{m^2 + p^2} + V(r).\eqno{(1.1)}$$ The eigenvalue equation
generated by this kind of Hamiltonian is usually called the
spinless Salpeter equation. It arises as a well-defined
approximation to the Bethe--Salpeter formalism for the description
of bound states within (relativistic) quantum field
theory~\cite{bse} when it is assumed that the bound-state
constituents interact instantaneously and propagate like free
particles~\cite{se}. At the same time, $H$ may be regarded~as the
simplest and perhaps most straightforward generalization of a
(nonrelativistic) Schr\"odinger operator towards the incorporation
of relativistic kinematics. For many potentials, this Hamiltonian
can be shown~\cite{lieb} to be bounded below and essentially
self-adjoint, and its spectrum can be defined variationally. For
definiteness, we consider the corresponding eigenvalue problem in
three spatial dimensions.

\section{The secant lower bound}The kinetic-energy operator $K =
\sqrt{m^2+p^2}$ is a concave function of the Schr\"odinger kinetic
energy $p^2.$ Hence, tangential operators to $K$ of the form
$\alpha + \beta p^2$ provide a class of Schr\"odinger upper bounds
to $K$. This idea has been explored and optimized in earlier
papers~\cite{halld,halle,hallf,hallsu}. In the present paper we are
concerned with lower bounds. The question arises as to whether any
of the family of Schr\"odinger operators $\alpha +\beta p^2 + V$
might generate a lower bound to $H$. On the basis of the usual
comparison theorem of quantum mechanics one would not expect this
since (in momentum space) the graph of $ \alpha + \beta k^2$
either lies above $K$ or crosses $K.$ However, under suitable
conditions, the comparison theorem has been
strengthened~\cite{hallct} to yield spectral inequalities even when
the corresponding potential graphs cross over. For our problem, we
must compare the two Hamiltonians $H = K + V$ and $H^{(s)} = m+
\beta p^2 + V,$ where $\alpha = m$ and $\beta >0$ is not yet
chosen; $V(r)$ is assumed to be a spherically symmetric attractive
potential in three spatial dimensions. Let us suppose that the
exact normalized ground state of $H$ is $\psi(r)$ and the
corresponding momentum-space function is $\phi(k)$: these are
normalized `radial' functions including a factor $r$ or $k$ and
satisfying, for example, $\psi(0) = 0,$
$\int_0^{\infty}\psi^2(r)dr = 1,$ and $$\phi(k) =
\left(\frac{2}{\pi}\right)^{\half}\int\limits_0^{\infty}
\sin(kr)\psi(r)dr.\eqno{(2.1)}$$Similarly, for the Schr\"odinger
comparison operator $H^{(s)}$, the wave functions are
$\psi^{(s)}(r)$ and $\phi^{(s)}(k).$ Following the same reasoning
as with two different potentials, which we used in the proof of
Theorem~3 in Ref.~[8], we consider the two eigenequations in
momentum space (where $V$ now becomes the integral operator
$\tilde{V}$): $$\left(\sqrt{m^2+k^2}+ \tilde{V}\right)\phi = E
\phi,\eqno{(2.2)}$$ $$\left(m + \beta k^2
+\tilde{V}\right)\phi^{(s)} = E^{(s)}\phi^{(s)}.\eqno{(2.3)}$$ If
we multiply (2.2) by $\phi^{(s)}$ and (2.3) by $\phi$, subtract,
and integrate on $[0,\infty)$, we obtain $$I =
\int\limits_0^{\infty}\left(\sqrt{m^2+k^2} - (m+\beta
k^2)\right)\phi(k)\phi^{(s)}(k)dk = \left(E-E^{(s)}\right)
\int\limits_0^{\infty}\phi(k)\phi^{(s)}(k)dk.\eqno{(2.4)}$$Now we
proceed to declare our assumptions and to choose $\beta.$ We first
define the function $W(k)$ as follows $$W(k) =
\int\limits_0^k\left(\sqrt{m^2+t^2} - (m+\beta
t^2)\right)\phi^{(s)}(t)tdt.\eqno{(2.5)}$$ The integral $I$ on the
left-hand side of (2.4) may then be integrated by parts to yield
$$I=-\int\limits_0^{\infty}W(k)\left(\frac{\phi(k)}{k}\right)^{\prime}dk.
\eqno{(2.6)}$$ We now show that $I \ge 0$ and that this in turn
proves that $E^{(s)} \le E.$ To this end we make some assumptions
concerning the two wave functions: (1) we assume that
$\phi^{(s)}(k) \ge 0$ and (2) that $\phi(k) \ge 0$ and also
$(\phi(k)/k)^{\prime} \le 0.$ That is to say, we assume that the
two wave functions are node free, and that the wave function for
the semirelativistic problem (with the factor $k$ removed) is
monotone non-increasing. These assumptions have to be considered
for each application. The final step is to choose $\beta $ so that
$I \ge 0.$ This is achieved by the requirement that $W(\infty) =
0.$ Clearly this determines $\beta.$ Moreover, the graphs of
$\sqrt{m^2+k^2}$ and $ m + \beta k^2$, which are shown in Fig.~1,
cross exactly twice, at $k = 0$, after which $K$ is immediately
larger than the Schr\"odinger operator, until they cross again.
Meanwhile the integral of the difference up to infinity is zero.
Thus we conclude $W(k) \ge 0.$ This combined with the assumed
positivity and monotonicity of $\phi(k)/k$ guarantees both that $I
\ge 0$ and that the integral on the right-hand side of (2.4) is
positive. Consequently, we have established the secant lower
bound, $E^{(s)} \le E.$ This completes the simple proof.

For nonrelativistic problems curious examples have been
constructed~\cite{hallcta} in which there are arbitrarily large
numbers of potential cross overs, but spectral ordering is still
guaranteed.

A natural application to consider would be the Coulomb problem
$V(r) = -c/r,$ with coupling not too large, that is, $c < 2/\pi.$
However, the integral in (2.5) is not defined for this problem
since the momentum-space expression of the exact Schr\"odinger
radial function is of the form $\phi^{(s)}(k) = Ak(a^2+k^2)^{-2},$
and one term also includes the factor $k^3.$ In the next section
we consider the example of the harmonic oscillator in some detail;
here there is no such difficulty since the momentum-space
Schr\"odinger wave function is Gaussian.
\begin{figure}[htbp]\centering\includegraphics[width=12cm]{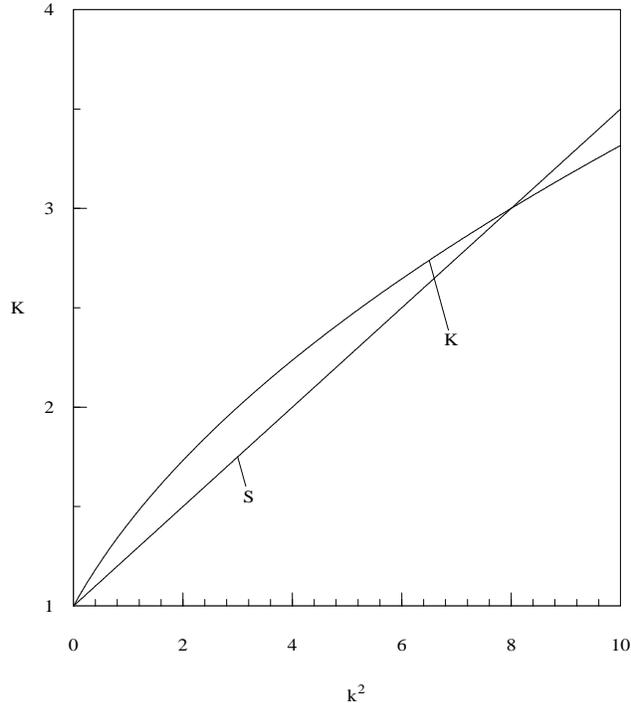}
\caption{Plot of semirelativistic $K = \sqrt{m^2 + k^2}$ and
Schr\"odinger $S = m + \beta k^2$ kinetic-energy functions against
$k^2$ in non-dimensional units with $m=1.$ The value $\beta =
0.2506$ is chosen so that $W(\infty) = 0$ for $V(r) = r^2,$ which
implies $W(k)\geq 0,\ k\geq 0.$} \label{Fig. (1)}\end{figure}

\section{An example}We now consider a test example for which there
are known (numerically) exact results. The harmonic oscillator is
equivalent to a nonrelativistic problem whose spectrum can be
determined numerically to high accuracy. Thus we have $$\sqrt{m^2
+ p^2} + r^2 \equiv p^2 + \sqrt{m^2 + r^2} \rightarrow
\epsilon_2(m),\eqno{(3.1)}$$ where $\epsilon_2(m)$ is the
ground-state energy of the semirelativistic oscillator in three
dimensions. Elementary scaling arguments then allow us to write
more generally, with coupling $c > 0,$ that $$\sqrt{m^2 + p^2} + c
r^2 \equiv c p^2 + \sqrt{m^2 + r^2} \rightarrow
c^{\frac{1}{3}}\epsilon_2(m c^{-\frac{1}{3}}).\eqno{(3.2)}$$
According to our present theory, a lower bound for this problem is
given by the Schr\"odinger operator $H^{(s)} = m + \beta p^2 + c
r^2$, with exact momentum-space eigenfunction $\phi^{(s)}(k).$ We
must first be sure that the unknown exact momentum-space wave
function $\phi(k)$ is node free, and that $\phi(k)/k$ is monotone
non-increasing. For well-behaved potentials, the ground state is
generally node free~\cite{lieb}. We know that the second condition
is also satisfied because of the following argument. In momentum
space the eigenequation for the semirelativistic problem may be
written $$-c\phi^{\prime\prime}(k) + \sqrt{m^2 +k^2}\ \phi(k) =
E\phi(k).\eqno{(3.3)}$$ The potential function $\sqrt{k^2+m^2}$ is
bounded below and is monotone increasing. Hence, by the result
proved at the start of Sec.~4 of Ref.~[8], we know that the
function $\phi(k)/k$ is indeed monotone non-increasing. We must
now choose $\beta$ to satisfy $W(\infty) = 0,$ that is to say
$$\int\limits_0^{\infty}\left(\sqrt{m^2+k^2}-\left(m + \beta
k^2\right)\right)\exp\left(-\half k^2 (\beta/c)^{\half}\right)k^2
dk = 0.\eqno{(3.4)}$$ After a change of variables and some
elementary Gaussian integrals, this condition may be written
$$g\left(\gamma^2\right) = \frac{\sqrt{\pi}}{2}\left(\gamma
+\frac{3m\beta}{2\gamma}\right),\eqno{(3.5)}$$ where the function
$g$ and the parameter $\gamma$ are defined by $$g(x) =
\int\limits_{-\infty}^{\infty}\left(x+t^2\right)^{\half}e^{-t^2}t^2
dt\quad {\rm and}\quad\gamma =
\left(\frac{m^4\beta}{4c}\right)^{\frac{1}{4}}.\eqno{(3.6)}$$
Thus, for each choice of coupling $c >0,$ the recipe for the lower
bound may be written $$m^3 =
\frac{6c\gamma^2}{\left(\frac{2}{\gamma\sqrt{\pi}}\right)g(\gamma^2)
-1}\rightarrow \boxit{0.03 true in}{$\gamma$}\ ,\quad\beta =
\frac{4c\gamma^4}{m^4},\quad E^L = m + 3\left(\beta
c\right)^{\half}.\eqno{(3.7)}$$ By taking the case $c = 1$ and
using these formulae, we find the results shown in Table~1.

\begin{table}[h]\caption{The secant lower bound $E^{(s)}$ and
corresponding accurate ground-state eigenvalues $E$ for the
problem $H=\sqrt{m^2+p^2}+r^2$ in ${\rm R}^3.$ The values of
$\beta$ are shown, which guarantee that the Schr\"odinger operator
$H^{(s)}=m+\beta p^2+r^2$, whose lowest energy is $E^{(s)},$
indeed provides a lower bound.}{\begin{tabular}{cccc}\hline\hline
$\qquad\qquad$$m$$\qquad\qquad$&$\qquad\qquad$$\beta$$\qquad\qquad$&
$\qquad\qquad$$E^{(s)}$$\qquad\qquad$&$\qquad\qquad$$E$$\qquad\qquad$\\\hline
0.1&0.4034&2.0055&2.3422\\0.2&0.3788&2.0464&2.3544\\
0.5&0.3190&2.1943&2.4323\\1&0.2506&2.5019&2.6640\\
2&0.1734&3.2492&3.3361\\3&0.1315&4.0880&4.1415\\
4&0.1056&4.9747&5.0105\\5&0.0879&5.8897&5.9153\\
7&0.0657&7.7692&7.7840\\10&0.0475&10.6539&10.6619\\
\hline\hline\end{tabular}}\end{table}

It is consistent with elementary physical arguments, and, indeed,
with the Schr\"odinger upper bounds discussed earlier~\cite{halle},
that the Schr\"odinger lower bounds presented here also show that
the semirelativistic problem becomes less relativistic as $m$
increases; in the limit $m\rightarrow\infty,$ both upper and lower
bounds approach the asymptotic form $$E\rightarrow m +
3\sqrt{\frac{c}{2m}}.$$

Similar results are obtained for the linear potential $V(r) = r.$
In this case we have $\psi^{(s)}(r) =
C\Ai(r\beta^{-\frac{1}{3}}-e_1),$ Where $\Ai$ is the Airy
function, and $e_1 \approx 2.33811$ is the bottom of the spectrum
of $p^2 + r.$ For $m = 2\sqrt{2},$ for example, we find $\beta
=0.13272$ and $E \geq 4.021,$ whereas an accurate numerical
value~\cite{bouk} is $E = 4.080.$
\vspace{0.4 in}

\section{Conclusion}Because of the concavity of the
semirelativistic Hamiltonian $H = \sqrt{m^2+p^2} + V(r)$ in $p^2,$
it would seem unlikely at first glance that one could find lower
bounds to the energy based on Schr\"odinger comparison operators.
In spite of this expectation, we show in this paper that such a
lower bound is possible. The secant bound involves a comparison
operator whose kinetic-energy function $m + \beta p^2$ has a graph
which crosses that of the semirelativistic expression $K =
\sqrt{m^2 + p^2}.$

\section*{Acknowledgements}Partial financial support of this work
under Grant No.~GP3438 from the Natural Sciences and Engineering
Research Council of Canada, and the hospitality of the Institute
for High Energy Physics of the Austrian Academy of Sciences in
Vienna, are gratefully acknowledged by one of us [RLH].

\end{document}